\documentstyle[12pt,aasms4]{article}

\newcommand\about     {\hbox{$\sim$}}

\def\r                {\hbox{$r^*$}}
\def\i                {\hbox{$i^*$}}

\def\ug               {\hbox{$u^*-g^*$}}
\def\gr               {\hbox{$g^*-r^*$}}
\def\ri               {\hbox{$r^*-i^*$}}
\def\iz               {\hbox{$i^*-z^*$}}
\def\mic              {\hbox{$\mu{\rm m}$}}
\def\refSDSSall       {\hbox{1}}
\def\refRADec         {\hbox{2}}
\def\reffilters       {\hbox{3}}
\def\refcmd           {\hbox{4}}
\def\refccd           {\hbox{5}}
\def\refcontours      {\hbox{6}}
\def\refdatamodels    {\hbox{7}}
\def\refhistograms    {\hbox{8}}
\def\Mo{\hbox{$M_{\odot}$}}

\begin{document}
  
\title{   Optical and Infrared Colors of Stars Observed by 
          the Two Micron All Sky Survey and 
          the Sloan Digital Sky Survey$^1$       }

\author{
Kristian Finlator\altaffilmark{\ref{Princeton}},
\v{Z}eljko Ivezi\'{c}\altaffilmark{\ref{Princeton}},
Xiaohui Fan\altaffilmark{\ref{Princeton}},
Michael A. Strauss\altaffilmark{\ref{Princeton}},
Gillian R. Knapp\altaffilmark{\ref{Princeton}}, 
Robert H. Lupton\altaffilmark{\ref{Princeton}},
James E. Gunn\altaffilmark{\ref{Princeton}},
Constance M. Rockosi\altaffilmark{\ref{Chicago}},
John E. Anderson\altaffilmark{\ref{Fermilab}}, 
Istv\'an Csabai\altaffilmark{\ref{JHU},\ref{Eotvos}},
Gregory S. Hennessy\altaffilmark{\ref{USNO}},
Robert B. Hindsley\altaffilmark{\ref{RSD}},
Timothy A. McKay\altaffilmark{\ref{Michigan}},
Robert C. Nichol\altaffilmark{\ref{CMU}},
Donald P. Schneider\altaffilmark{\ref{PennState}},
J. Allyn Smith\altaffilmark{\ref{Michigan}},
Donald G. York\altaffilmark{\ref{Chicago}}
for the SDSS Collaboration
}

\altaffiltext{1}{Based on observations obtained with the
Sloan Digital Sky Survey.}
\newcounter{address}
\setcounter{address}{2}
\altaffiltext{\theaddress}{Princeton University Observatory, Princeton, NJ 08544
\label{Princeton}}
\addtocounter{address}{1}
\altaffiltext{\theaddress}{University of Chicago, Astronomy \& Astrophysics
Center, 5640 S. Ellis Ave., Chicago, IL 60637
\label{Chicago}}
\addtocounter{address}{1}
\altaffiltext{\theaddress}{Fermi National Accelerator Laboratory, P.O. Box 500,
Batavia, IL 60510
\label{Fermilab}}
\addtocounter{address}{1}
\altaffiltext{\theaddress}{Dept. of Physics and Astronomy, 
The Johns Hopkins University, 3701 San Martin Drive, Baltimore, MD 21218
\label{JHU}}
\addtocounter{address}{1}
\altaffiltext{\theaddress}{Dept. of Physics of Complex Systems,
E\"otv\"os University, P\'azm\'any P\'eter s\'et\'any 1/A, Budapest, H-1117, Hungary
\label{Eotvos}}
\addtocounter{address}{1}
\altaffiltext{\theaddress}{U.S. Naval Observatory,
3450 Massachusetts Ave., NW,
Washington, DC  20392-5420
\label{USNO}}
\addtocounter{address}{1}
\altaffiltext{\theaddress}{Remote Sensing Division, Code 7215, Naval Research 
Laboratory, 4555, Overlook Ave. SW, Washington, DC 20375
\label{RSD}}
\addtocounter{address}{1}
\altaffiltext{\theaddress}{University of Michigan, Dept. of Physics,
500 East University, Ann Arbor, MI 48109
\label{Michigan}}
\addtocounter{address}{1}
\altaffiltext{\theaddress}{Dept. of Physics, Carnegie Mellon University, 
5000 Forbes Ave., Pittsburgh, PA 15232
\label{CMU}}
\addtocounter{address}{1}
\altaffiltext{\theaddress}{Dept. of Astronomy and Astrophysics,
The Pennsylvania State University,
University Park, PA 16802
\label{PennState}}

\begin{abstract}     

We discuss optical and infrared photometric properties of stars matched
in the Two Micron All Sky Survey (2MASS) and the Sloan Digital Sky Survey
(SDSS) commissioning data for $\about$50 deg$^2$ of sky along
the Celestial Equator centered at $l=150^\circ, b=-60^\circ$. 
About 98\% ($\about$63,000) of objects listed in the 2MASS 
Point Source Catalog in the analyzed area are matched within 2 arcsec to an 
SDSS source. The matched sources represent 8\% of the $\about$800,000 SDSS 
sources in this area. They are predominantly red sources, as expected, 
and 15\% of them are resolved in SDSS imaging data, although they are 
detected as point sources in 2MASS data.
The distribution of positional discrepancies for the matched sources, 
and the astrometric statistics for the multiply observed SDSS sources, 
imply that the astrometric accuracy of both surveys is about 0.1 arcsec 
per coordinate (rms). 

For about 14,000 stars with the smallest photometric errors ($\la$ 10\%) in
both surveys, we present optical and infrared color-magnitude and color-color 
diagrams. We use optical (SDSS) colors to identify the stellar spectral sequence
and show that stars of different spectral types can have similar infrared colors,
thus making the classification of stars based on only 2MASS data very 
difficult. However, a broad separation into ``early'' and ``late'' spectral 
types (relative to type K0) is possible with a reliability of $\about$95\% even 
with 2MASS colors alone. 

The distributions of matched sources in color-magnitude and color-color diagrams 
are compared to the predictions of a stellar population synthesis code. We find that 
the models are in fair overall agreement with the data. In particular, the total 
number counts agree to better than 10\%, and the morphologies of the color-magnitude 
and color-color diagrams appear similar. The most significant discrepancies are 
found for the number ratio of ``early'' to ``late'' type stars (by about a factor of 
2) and in the colors of M stars (up to 0.2 mag). The first disagreement indicates 
that some parameters of the standard Galactic structure model and/or initial mass
function can be improved, and the second disagreement emphasizes known difficulties 
with the modeling of stellar atmospheres for cool stars.

\end{abstract}


\section{             Introduction              }

The Two Micron All Sky Survey (2MASS, Skrutskie {\em et al.} 1997) and the Sloan 
Digital Sky Survey (SDSS, York {\em et al.} 2000) are producing eight-color optical 
and infrared photometry for millions of Galactic and extragalactic sources. 
It is of obvious scientific interest to positionally match the sources
observed in both surveys and to study their distribution in color-magnitude 
and color-color diagrams. Due to the large number of sources
and very accurate photometry and astrometry, such a study can, for example, 
reliably reveal detailed properties of ``normal" stars which should constitute the 
majority of matched sources. This information is invaluable 
for studying stellar and Galactic structure and evolution, and also makes 
feasible efficient searches for peculiar, or previously unknown objects. For
example, the selection of recently defined L dwarfs (Kirkpatrick {\em et
al.} 1999) was based on both optical and infrared photometry, an approach still 
followed in the current searches for L and T dwarfs (e.g. Leggett {\em et al.} 
2000).

In this paper we present preliminary results for matched sources from the recent 
2MASS Second Incremental Data 
Release\footnote{http://www.ipac.caltech.edu/2mass/releases/second/index.html}
and SDSS commissioning data. A brief description of both surveys is provided in \S 2. 
We describe the matching and the basic statistics in \S 3. A more detailed study 
of optical-infrared properties of stars is presented in \S 4, and in \S 5 we compare 
data with models. 

\section {  The Data } 

\subsection {2MASS} 

The 2MASS is using two 1.3-meter telescopes, one at 
Mt. Hopkins, AZ, and one at CTIO, Chile, to survey the entire sky in
near-infrared light\footnote{http://www.ipac.caltech.edu/2mass/overview/about2mass.html}.  
The Northern telescope has been observing regularly since June 1997, while 
the Southern facility began operation in March 1998. Each telescope's camera is 
equipped with three $256\times256$ arrays (the pixel size is 2 arcsec) of HgCdTe 
detectors and observes simultaneously in the $J$ (1.25 $\mic$), $H$ (1.65 $\mic$), 
and $K_s$ (2.17 $\mic$) bands.  The detectors are sensitive to point sources brighter 
than about 1 mJy at the $10\sigma$ level, corresponding to limiting (Vega-based) magnitudes 
of 15.8, 15.1, and 14.3, respectively.  Point-source photometry is repeatable to
better than 10\% precision at this level, and the astrometric uncertainty for these 
sources is less than 0.2 arcsec.

When completed, 2MASS will catalog $\sim$300 million stars as well as several million
galaxies, and will cover at least 95\% of the sky.  A new data installment, the 
2MASS Second Incremental Data Release, was recently made available to the public.  
Spanning 47\% of the sky, it contains photometry and astrometry for over 162 million 
point sources as well as 1.9 million resolved sources.

\subsection {SDSS}

The SDSS is a digital photometric and spectroscopic survey which will 
cover one quarter of the Celestial Sphere in the North Galactic cap and produce
a smaller area ($\sim$ 225 deg$^2$) but much deeper survey in the Southern 
Galactic hemisphere (York {\em et al.} 2000\footnote{see 
also http://www.astro.princeton.edu/PBOOK/welcome.htm}). The flux densities of
detected objects are measured almost simultaneously in five bands ($u'$, $g'$, $r'$, 
$i'$, and $z'$; \cite{F96}) with effective wavelengths of 3540~\AA, 
4760~\AA, 6280~\AA, 7690~\AA, and 9250~\AA, 95\% complete\footnote{These 
values are determined by comparing multiple scans of the same area obtained 
during the commissioning year. Typical seeing in these observations was 
1.5$\pm$0.1 arcsec.} for point sources to limiting magnitudes of 22.1, 22.4, 22.1, 21.2,
and 20.3 in the North Galactic cap\footnote{We refer to the measured magnitudes in 
this paper as $u^*, g^*, r^*, i^*,$ and $z^*$ because the absolute calibration 
of the SDSS photometric system is still uncertain at the $\sim 0.03^m$ level. 
The SDSS filters themselves are referred to as $u', g', r', i',$ and $z'$. All 
magnitudes are given on the AB$_\nu$ system (Oke \& Gunn 1983, for additional 
discussion regarding the SDSS photometric system see \cite{F96} and Fan 1999).}. 
The survey sky coverage of about $\pi$ steradians (10,000 deg$^2$) will result 
in photometric measurements to the above detection limits for about 50 million stars. 
Astrometric positions are accurate to about 0.1 arcsec per coordinate for sources 
brighter than $r^* \sim$ 20.5$^m$, and the morphological information from the 
images allows robust  star-galaxy separation to $r^* \sim$ 21.5$^m$. 

\section         {          Matched Data                      }

\subsection {              SDSS Data     }

We utilize a portion of SDSS imaging data from two commissioning runs 
(numbered 94 and 125) taken during the Fall of 1998. The data in each run 
were obtained in six parallel scanlines\footnote{See also 
http://www.astro.princeton.edu/PBOOK/strategy/strategy.htm}, 
each 13.5 arcmin wide, along the Celestial Equator 
($-1.2687^\circ <$ $\delta_{J2000}$ $< 1.2676^\circ$). The six scanlines from each run are
interleaved to make a filled stripe. The portion of data used in this 
work extends from $\alpha_{J2000}$ = 0$^h$ 24$^m$ ($l=108^\circ$, $b=-62^\circ$) to 
$\alpha_{J2000}$ = 3$^h$ 0$^m$ ($l$=177$^\circ$, $b=-49^\circ$). 
The seeing in both runs was variable between 1 and 2 arcsec 
(FWHM) with the median value typically 1.5 arcsec. The total area which overlaps 
with public 2MASS data is 47.41 deg$^2$ and includes about 800,000 SDSS sources 
to the $6\sigma$ detection limit in at least one of the SDSS bands.

The SDSS color-color and color-magnitude diagrams which summarize photometric
properties of unresolved sources are shown in Figure \refSDSSall. In this and
all other figures, we correct the data for the interstellar 
extinction\footnote{The Galactic structure model discussed in \S5 implies
that the most sources in this sample are further than 200 pc away, which
justifies this correction.}   
determined from the maps given by Schlegel, Finkbeiner \& Davis (1998). 
Typical values for high-latitude regions discussed in this work are  
$A_{r^*}$ = 0.05-0.15 mag ($A_{r^*} = 0.84 A_V$). Throughout this 
work we use the ``point spread function'' (PSF) magnitudes\footnote{Note that
SDSS photometric system uses asinh magnitudes (\cite{LGS99}). For reference, 
zero flux corresponds to asinh magnitudes of 23.40, 24.22, 23.98, 23.51, and 
21.83 in $u^*, g^*, r^*, i^*$, and $z^*$, respectively.}
(measured by fitting a PSF model of a double Gaussian and applying an aperture
correction) as computed by the photometric pipeline (``photo'', version v5\_0; 
for details see \cite{Lupton00}). Photometric errors are typically 0.03$^m$ at the 
bright end ($r^* < 20^m$), relevant in this work (for more details see 
Ivezi\'c {\em et al.} 2000). 

The top left panel of Figure 1 displays the \r\ vs. \gr\ color-magnitude 
diagram for $\about$ 25,000 objects observed in 3 deg$^2$ of sky during SDSS 
commissioning run 94. There are at least three distinct groups of sources. 
The red sources with \gr\ $\about$ 1.4 are disk stars, and are dominated by 
stars of spectral type M, see e.g. Fukugita {\em et al.} (1996) and Lenz 
{\em et al.} (1998). The blue ridge with \gr\ $\about$ 0.4 is dominated by 
F and G disk stars at the bright end (\r $\la$ 18), and by halo stars at the 
faint end (note that the median \gr\ color for halo stars is bluer than for disk 
stars, due to the lower metallicity in the halo).  As shown by Fan (1999),
the contribution of low-redshift ($z \la 2$) quasars (QSOs) becomes increasingly 
important at the faint end of the blue ridge. The majority of stars observed by
SDSS are on main sequence; with the aid of models discussed in \S5 we estimate
that the fraction of stars which are not on main sequence is $\sim$1.0\%. 
The latter are dominated by giants and subgiants with 0.4 $<$ \gr $<$ 0.8 (\about 
90\%) and horizontal branch stars with \gr $<$ 0.5 (\about 10\%).

The three remaining panels show three projections of the SDSS color-color diagrams 
for objects, marked as dots, brighter than 20$^m$ in each of the 3 bands used to 
construct each diagram. Red is always towards the upper right corner. Objects that 
have \ug\ and \gr\ colors similar to low-redshift QSOs (\ug $<$ 0.4, -0.1 $<$ \gr $<$ 
0.3, \ri $<$ 0.5), and are brighter than the limit for quasars in the SDSS 
spectroscopic survey (\i $<$ 19), are marked by open circles. 

The locus of ``normal" stars stands out in all three diagrams. 
As shown by Fukugita {\em et al.} (1996), Krisciunas, Margon \&
Szkody (1998), Lenz {\em et al.} (1998), and Fan (1999), the position of a star 
in these diagrams is mainly determined by its spectral type. For reference, 
the positions of several spectral types taken from Fukugita {\em et al.} (1996) 
are marked next to the stellar locus in the \ug\ vs. \gr\ and \ri\ vs. \gr\ color-color 
diagrams (labels are slightly offset for clarity). 

For most of its length, the locus in the \gr\ vs. \ug\ diagram is made of stars with 
spectral types ranging from F to late K. A-type stars are visible at \ug $\about$ 1.0--1.2 
and \gr $\about$ -0.3--0.0 (for a detailed study of A stars detected in SDSS commissioning
data see Yanny {\em et al.} 2000), and late K and M stars are found 
at the red end of the stellar locus. Different M spectral subtypes cannot be 
distinguished in the \gr\ vs. \ug\ diagram, and these stars are better separated in the 
\ri\ vs. \gr\ diagram. M stars have a constant \gr\ color (\about 1.4) due to strong 
TiO bands in their spectra, and a value of \ri\ color that depends strongly on spectral
subtype. In the \iz\ vs. \ri\ diagram, stars earlier than K type 
are found in a small region around \ri $\about$ 0.2 and \iz $\about$ 0.0. 
For most of its length, the \iz\ vs. \ri\ stellar locus represents late K and M
stars. 

\subsection      {         2MASS       Data          }
 
We positionally match data from the two SDSS commissioning runs to 2MASS 
Point Source Catalog (PSC) data with 
0$^h$ 24$^m$ $<$ $\alpha_{J2000}$ $<$ 3$^h$ 0$^m$ and 
$-1.2687^\circ <$ $\delta_{J2000}$ $< 0^\circ$ (2MASS data for 
0$^\circ$ $<$ $\delta_{J2000}$ $< 1.2676^\circ$ are not yet available). The region
with 1$^h$ 51$^m$ $<$ $\alpha_{J2000}$ $<$ 1$^h$ 57$^m$ is missing from the 2MASS catalog.
The resulting overlapping area (47.41 deg$^2$) includes 64,695 2MASS sources.

\subsection   { The Positional Matching of SDSS and 2MASS Catalogs }

We consider as matches all sources from both catalogs whose listed positions 
agree to better than 2 arcsec. There are 63,341 such sources, or 97.9\% of the 
2MASS PSC sources and 8\% of the SDSS sources, in the overlapping area.
The probability of finding an SDSS source within a circle with 2 arcsec 
radius is \about 1\%, and the probability of associating a 2MASS source
with multiple SDSS sources is thus negligible ($<$ 10$^{-4}$). With the 
further constraints on the final sample, discussed in \S4, the random 
association rate for matched sources also becomes inconsequential 
($<$ 10$^{-5}$).

The distributions of RA and Dec differences between the SDSS and 2MASS 
catalog positions are shown in Figure \refRADec. The thin lines 
show the normalized histograms for all matched sources, and the thick lines 
show histograms for the $\about 14,000$ sources with at least 10$\sigma$ detections 
in both surveys (see the next section), which are not saturated or blended with 
other sources in the SDSS images. Both histograms are normalized by the total
number of sources. Equivalent Gaussian widths determined from the
interquartile range are 0.30 arcsec for the distributions of all sources and
0.15 arcsec for the sources with the high signal-to-noise ratio\footnote{The 
contribution of the proper motions to this scatter is negligible due to 
the short time span between the SDSS and 2MASS observations.}  (SNR). As the
multiple SDSS commissioning observations of the same area show that the 
positions of sources with high SNR are reproducible to 0.10 arcsec per 
coordinate (Pier {\em et al.} 2000), 
this implies that the astrometric accuracy of both 2MASS and SDSS is about 
0.10 arcsec for bright sources (positional root-mean-square scatter per 
coordinate). However, note that the histograms displayed in Figure \refRADec\ 
show a 0.05 arcsec offset from the origin for both coordinates. 
A comparison of the SDSS and VLA FIRST survey (Faint Images of the Radio Sky 
at Twenty-cm, Becker {\em et al.} 1995) data shows the same offset (though with 
less statistical significance due to the smaller number of matched sources), 
indicating that this is a systematic error in the SDSS astrometry. We are currently
investigating this problem.

\subsection     {      Unmatched  2MASS PSC  Sources        }
 
There are 1354 2MASS sources (2.1\% of the whole sample) which do not have
a cataloged SDSS source within 2 arcsec. Visual inspection of SDSS images 
shows that 70\% of these do not have an optical counterpart at the cataloged 
2MASS position. Such sources are either spurious 2MASS detections (roughly 80\% of 
the unmatched 2MASS sources have SNR $<$ 10 in all 2MASS bands), extremely
red sources (e.g. stars heavily obscured by circumstellar dust), or asteroids 
which moved between the times of the SDSS and 2MASS observations. 
There is no significant correlation between the number of unmatched sources
and the distance from the Ecliptic Plane.  

15\% of the unmatched 2MASS sources are associated in SDSS images with saturated 
stars. The centroids of saturated stars in SDSS have a large positional uncertainty 
which sometimes may exceed the matching cutoff (2 arcsec). 10\% of the unmatched 
2MASS sources are associated in SDSS images with complex structures consisting 
of several blended sources, and which were not deblended by the processing 
software.  The remaining 5\% of the unmatched sources are blended with a 
satellite trail or a diffraction spike of a nearby bright star in the SDSS data. 
This category represents only 0.1\% of the total number of 2MASS sources.

\section {     Optical and Infrared Properties of Stellar Sources }

The combined SDSS and 2MASS eight-color photometry can be used to construct
a multitude of point source color-magnitude and color-color diagrams. The five SDSS 
and three 2MASS bandpasses are shown in Figure \reffilters. All curves include 
typical atmospheric extinction and the transmission of the entire instrumental system.
The solid curves correspond to the overall efficiencies of the SDSS system 
(Gunn {\em et al.} 1998), and the dot-dashed lines show the 2MASS responses
normalized to 0.5 (M. Cohen, priv. comm.).

Since the thrust of this work is to provide the accurate optical and infrared 
colors of normal stars, we apply several conditions on the matched
sources in order to select a subsample of stars with the best photometric 
data.

\begin{enumerate}

\item 
The 2MASS and SDSS positions must agree to better than 1 arcsec. This 
limit corresponds to a $\about$5$\sigma$ cut.

\item 
Sources must not be saturated in any band in the SDSS data. This is roughly equivalent 
to the condition $i^* > 14$, because the majority of matched sources are red. 

\item
Sources must have high-SNR 2MASS detections. We impose 
this limit by requiring $J < 15.8$, $H < 15.1$ and $K_s < 14.3$, which corresponds
to $10\sigma$ detections. Since the majority of matched sources have $J-K_s$ colors 
bluer than 1.5 (see below), the faint limit is in practice determined by the 
condition on $K_s$ flux.

\item 
Sources must be unresolved in SDSS data. The SDSS star-galaxy separation is
robust to $r^* \about 21.5$, which is significantly fainter than the faint 
limit of the resulting sample, and only a negligible number
of galaxies are likely to be present in the final sample.  

\end{enumerate}

These limits result in a sample of 13,924 sources. Note that the fourth condition 
does not necessarily exclude all extragalactic sources since QSOs are typically 
optically unresolved sources. However, low-redshift QSOs can be easily recognized 
in the SDSS \gr\ vs. \ug\ color-color diagram (c.f. Figure 1) as shown by
Fan (1999). There are 100 matched sources in this region of the color-color diagram,
but only seven of these are bright enough to satisfy the (conservative) 2MASS 
SNR limits (\# 3 above). These sources, together with the optically classified galaxies, 
will be discussed in a separate publication.

Figure \refcmd\ displays color-magnitude diagrams for the 13,917 selected 
stellar sources. The SDSS and 2MASS photometry is better than 10\% for the 
majority of this sample due to the imposed cuts. The upper left panel shows
an \r\ vs. \gr\ diagram and can be compared to the analogous diagram for all
SDSS sources displayed in the upper left panel in Figure \refSDSSall. The 
bright limit on the \r\ magnitude is correlated with the \gr\ color because
the red sources become saturated in SDSS data at a fainter \r\ magnitude than
the blue sources. Similarly, the faint limit depends even more strongly on
the \gr\ color because only comparatively bright blue objects can pass 
the 2MASS faint limit cutoffs due to their steeply falling spectral
energy distributions\footnote{Due to these cutoffs, essentially all matched
stars belong to the Galactic disk. With the aid of models discussed in \S5, we 
estimate that $\sim$97\% of these stars are on main sequence.}.
Essentially the same effects can be seen in the other
three color-magnitude diagrams which combine optical and infrared data.
Note the sharp faint limit in the $J$ vs. $J-K_s$ diagram because it is
determined by the $K_s < 14.3$ condition, as discussed above.

The color-color diagrams for the selected subsample are shown in
Figure \refccd. The upper left panel shows an analogous diagram to the
one displayed in the lower left panel of Figure \refSDSSall. Note that
there is no significant difference in the appearance of the stellar 
locus. However, as is already visible in Figure \refcmd, the matched
sample has a larger fraction of red stars (see below). The following
three panels show optical-infrared color-color diagrams. Note that the
stellar locus is clearly visible in all of them, confirming that the
optical and near-infrared
broad-band colors of normal stars can be described to within 0.1-0.2 
magnitudes in terms of temperature as a single parameter.
In particular, the near-infrared fluxes can be predicted for the majority of 
sources to 0.1-0.2 mag from the optical fluxes alone. However, 
the converse is not true; stars with varying optical 
colors may occupy the same region in near-infrared color-color diagrams
(n.b. in the Rayleigh-Jeans tail color is independent of temperature).
This is true for some stars even in the case of optical-infrared color-color
diagrams. For example, all M stars occupy a very small region in the
$J-K_s$ vs. \gr\ diagram shown in the middle left panel (\gr = 1.4, 
$J-K_s$=0.85). While their \gr\ color can be determined from their
$J-K_s$ color, it is impossible, for example, to constrain 
their \ri\ color, which can vary by more than a magnitude. 

The lower right panel displays the $J-K_s$ vs. $r^*-K_s$ color-color diagram. 
This diagram was used by Kirkpatrick {\em et al.} (1999) in their search for 
L dwarfs; we present it here as reference (note that Kirkpatrick {\em et al.} 
display this graph with reversed $J-K_s$ axes).
Kirkpatrick {\em et al.} show that the M spectral sequence begins around
$r^*-K_s$ = 2.0, and ends around $r^*-K_s = 8.0$, where the L dwarf sequence
begins. L dwarfs are very rare and are not apparent on the diagram.
The position of Gl 229B, the first discovered T dwarf (Nakajima 
{\em et al.} 1995), is marked by an open circle at $r^*-K_s = 9.1$ and
$J-K_s = -0.1$. 

The lower left panel shows a color-color diagram constructed from 
2MASS data alone. The stellar locus is not as well defined as in the optical 
and optical-infrared color-color diagrams, and the position of a star in 
this diagram is not strongly correlated with its spectral type. 
However, the structure seen in the $H-K_s$ vs. $J-H$ color-color diagram 
indicates that there is at least a weak correlation between the spectral type 
and the position in this diagram. Understanding such a correlation could 
allow for very efficient studies of the Galactic disk structure based on only 
2MASS data; unlike SDSS, 2MASS will extend over the entire sky.
We attempt to quantify such a correlation by separating sources based on their 
optical colors, and then studying their distribution in the $H-K_s$ vs. $J-H$
color-color diagram.

The \ri\ vs. \gr\ color-color diagram can be used to separate 
the stars into rough spectral classes (c.f. \S 3.1). We divide the sample into

\begin{enumerate}
\item
Stars with spectral types earlier than G5 by the condition $\gr \le 0.5$.
There are 2894 stars (21\%) in this subsample.
\item
Stars with spectral types between G5 and K5: $0.5 < \gr \le 1.0$.
There are 3731 stars (27\%) in this subsample.
\item 
Stars with spectral types between K5 and M5: $\gr > 1.0$ and
$\ri \le 1.0$. There are 4393 stars (32\%) in this subsample.
\item 
Stars with spectral types later than M5: $\gr > 1.0$ and
$\ri > 1.0$. There are 2873 stars (20\%) in this subsample.

\end{enumerate}

The distribution of stars in the $H-K_s$ vs. $J-H$ and $J-K_s$ vs. $r^*-K_s$ color-color
diagrams is shown separately for each subsample in the four panels of Figure \refcontours. 
The spectral types are plotted in the same order as listed above, with stars earlier
than type G5 shown in the top panel. Stars from each optically selected spectral subsample
are plotted as dots, and the contours represent the distribution of the whole sample.
The contour levels are 0.5, 0.3, 0.1 and 0.03 of the peak density. 
The separation of the four classes is much better defined in the $J-K_s$ vs. $r^*-K_s$ 
diagram, because the range of colors is much larger than in the $H-K_s$ vs. $J-H$ diagram.
The two peaks in the $H-K_s$ vs. $J-H$ source distribution correspond to stars earlier 
than G5 (blue peak), and to stars with spectral types later than K5 (red peak). Stars 
with spectral types between G5 and K5 populate the area between the peaks, and the 
stars with spectral types later than M5 populate the region with the reddest $J-K_s$ 
and $r^*-K_s$ colors. 

We have also attempted a finer binning in the \ri\ vs. \gr\ color-color diagram and found 
that the 2MASS data alone does not allow for a more precise spectral classification than that
presented in Figure \refcontours. Nevertheless, we find that the 2MASS color $J-K_s = 0.6$ 
provides a good overall separation of stars with $\gr < 0.7$ from stars with $\gr \ge 0.7$
(c.f. Figure 5). There are 5099 stars (37\%) with $\gr < 0.7$ in the subsample of 13917 
stars with good photometry, and 93\% of these also have $J-K_s < 0.6$. Of the remaining 8818 
stars with $\gr \ge 0.7$, there are 96\% with $J-K_s \ge 0.6$. These numbers show that
there is a $\about95\%$ probability of success in predicting whether the SDSS \gr\ color 
is greater or smaller than 0.7 from the information specifying whether the 2MASS $J-K_s$ color  
is greater or smaller than 0.6. The SDSS \gr\ color of 0.7 separates stars with 
spectral type earlier than K0 from stars with later spectral types. This boundary 
is fuzzy to within 2-3 spectral subtypes due to the effects of varying metallicity and 
gravity on stellar colors. Such a method of separating early from late type stars may 
prove useful in selecting subsamples of 2MASS data for the studies of Galactic disk 
structure.

\section { Comparison of Data with a Stellar Population Synthesis Code  }

We compare the number counts and the source distribution in color-color 
diagrams using the code of Fan (1999, hereafter F99) for simulating SDSS observations 
of unresolved sources, which we have extended to the 2MASS bands (Finlator 2000).
F99 compared several simulated stellar populations to data obtained by the ESO Imaging
Survey Patch B (Prandoni {\em et al.} 1998), and found that the position of the 
simulated stellar locus in SDSS color-color space agreed very well with the data.
Nevertheless, he also found that the number counts could differ by up to $\sim$20\%. 
In the following section we present an expanded analysis of the model predictions,
made possible by the new 2MASS and SDSS data, and discuss similar discrepancies.
Detailed discussion of this model is given by F99, and here we only provide a brief 
summary.

\subsection{       Generating a Simulated Population      }

The population synthesis code for generating a simulated stellar population at 
a given Galactic latitude and longitude is based on the following inputs/steps:

\begin{enumerate}

\item
A Galactic structure model based on the star count models of Bahcall \& Soneira 
(1981, 1984) is used to generate the number of halo and disk stars as a function of 
Galactic latitude and longitude.  The disk scale height is assumed to increase
with stellar age (i.e. the scale height is correlated with the spectral type), 
and the Galactic halo is flattened. The initial mass function is taken from 
Haywood {\em et al.} (1997) and is assumed to be valid both for disk and halo stars. 

\item 
To generate the properties of a specified stellar population, the simulation
generates the mass, age, and metallicity of each star with the aid of the Padova 
stellar evolutionary tracks and isochrones (Bertelli et al. 1994). The
resulting $T_{eff}$, [Fe/H], and log($g$) are then used to select a stellar
atmosphere model in order to determine the colors for each star.

\item
The hybrid model stellar atmospheres compiled by Lejeune, Buser, \&
Cuisinier (1997a, 1997b) are used to compute the SDSS and 2MASS colors of 
stars, given their effective temperature $T_{eff}$, metallicity [Fe/H], and 
surface gravity log($g$). 
 
\item
The SDSS AB magnitudes are computed as described in F99, and the 2MASS
Vega magnitudes, $m$, are computed using
\begin{equation}
   m = -2.5 \log \frac{\int \mathrm{d}(\log(\nu))f_{\nu}S_{\nu}}
            {\int \mathrm{d}(\log(\nu))S_{\nu}} - 48.60 + m_0,
\end{equation}
where $m_0(J)=-0.89$, $m_0(H)=-1.37$, and $m_0(K_s)=-1.84$ represent the offsets
from the AB magnitude system. These offsets
were computed  by assuming that the flux from Vega is constant within any 2MASS 
band and equal to 1595, 1024, and 667 Jy in the $J$, $H$, and $K_s$ bands, 
respectively (M. Cohen, priv. comm.).

\item
Photometric errors in SDSS bands are added as described in Appendix A of F99.  
The 2MASS photometric errors are derived from an online plot of photometric 
repeatability versus
magnitude\footnote{http://www.ipac.caltech.edu/2mass/releases/second/doc/figures/secii2f7.gif}.
Errors are $\about0.03^m$ at the bright end, and then rise to $\about0.1^m$ at
$J=15.8$, $H=15.1$, $K_s=14.3$.

\end{enumerate}

\subsection {Comparison of Simulated Observations with the Data   }

Figure \refdatamodels\ compares the photometric properties of
the matched sources with a population simulated at 
$(l,b)=(150^\circ,-60^\circ)$ (the 
properties of a sample are strong functions of the Galactic coordinates)  
and spanning the same area on the sky (47.41 deg$^2$). The panels on the left 
show the diagrams for the matched objects, and those on the right show the 
simulated population. To ensure the same selection effects for both 
samples, we require that the model fluxes satisfy $z^* > 14$, $J < 15.8$,
$H < 15.1$, and $K_s < 14.3$. These conditions result in 12,740 simulated 
sources, only 9\% fewer than the observed value of 13,917 matched sources. 
Since the selection effects are quite complicated, and include two independent 
surveys, this close agreement is quite encouraging. We note that the simulated 
population does not include ``exotic'' objects such as white dwarfs and QSOs.

The top row in Figure \refdatamodels\ compares optical colors.
The overall agreement is satisfactory, although
there are fewer simulated stars with \ri $\about$ 0.8-0.9, yielding a 
discontinuity in the stellar locus in the \gr\ vs. \ri\ diagram.
This is due to the different model libraries used for objects with T$_{eff} >$ 
3500 K and T$_{eff} <$ 3500 K, as discussed by F99. 
The middle row in Figure \refdatamodels\ shows an optical-infrared color-color 
diagram. The overall morphology is well reproduced by the simulated 
population, except for the M stars ($r^*-K_s \ga 3$) which have notably 
different colors. While the observations suggest that infrared colors should 
not vary with $r^*-K_s$ color for $r^*-K_s \ga 3$, the simulated M stars 
become bluer in $J-K_s$ after this point. We will further discuss this discrepancy 
below. The bottom row in Figure \refdatamodels\ compares the $z^*$ vs. $r^*-K_s$ 
color-magnitude diagrams. Again, the overall morphology is reproduced, but the 
simulation does not seem to generate enough bright stars with intermediate 
$r^*-K_s$ colors. 

To provide a more quantitative comparison of the data with the models, 
we display histograms of the matched and simulated objects in Figure 
\refhistograms. The top four plots show that the simulation
overestimates the overall counts of blue stars while underestimating the
red star counts. While the observed ratio of stars redder and bluer than 
$J - K_s = 0.6$ is 1.73, the simulated ratio is 0.98.
The top right panel clearly shows the lack of simulated stars with \ri $\about$ 
0.8-0.9 mentioned earlier.

The disagreement between the observed and simulated ratio of red to blue stars 
could be due either to incorrect parameters of the Galactic structure model, and
to an incorrect shape of the initial mass function employed in the simulation. 
The blue stars have higher luminosities than do the red stars and thus are observed 
to larger distances. The majority of simulated red stars with $J - K_s > 0.6$ 
have distances ranging from 200 to 800 pc (median value of 500 pc), while 
the majority of blue stars are in the distance range 800-2000 pc (median
value of 1000 pc). Decreasing the disk scale height for red stars, and
increasing it for blue stars in the expression for stellar density as a function 
of distance from the Galactic plane (F99, eqs. 5 and 6) can increase the simulated
ratio of red to blue stars. However, these adjustments would involve changing the 
age of stars,  resulting in a change in the total number of simulated stars. 

Similarly, one can increase the simulated ratio of red to blue stars by increasing
the slope of the initial mass function (IMF) at the low-mass end. 
The simulated sample is dominated by stars with masses $\la 1 \Mo$. The majority of 
stars with $J - K_s > 0.6$ have masses between 0.3 \Mo\ and 0.75 \Mo, while the 
stars with $J - K_s < 0.6$ have masses between 0.75 \Mo\ and 1.1 \Mo. In this
mass range, the simulation adopts $dN/dM \propto M^{-1.7}$. We have examined a set
of models with $dN/dM \propto M^{-n}$ and found that $n=2.5\pm0.1$ produces
the observed ratio of red to blue stars. However, the total
number of stars becomes too low and another model adjustment is needed.
Thus, to meaningfully constrain the free parameters requires an extensive modeling 
effort beyond the scope of this work. A detailed analysis of SDSS stars counts
and their implications for the Galactic structure model will be reported in
Chen {\em et al.} (2000).  

Another model difficulty is related to the properties of the stellar models, 
as opposed to the Galactic distribution of stars. As discussed above, the $J-K_s$ colors 
of simulated M stars becomes bluer for $r^*-K_s \ga 3$, while the observations suggest 
that they should stay constant at roughly 0.85. The bottom two panels of Figure 8
show the $J-K_s$
histograms for two subsamples of M stars: the left panel shows stars with
$3.5 < \r -K_s < 4$ ($\about$ M3) and the right panel shows stars with
$4.5 < \r -K_s < 5$ ($\about$ M5). It is evident that the $J-K_s$ color
for simulated early M stars is significantly redder (0.15$^m$) than observed,
while it is bluer than observed for the mid-to-late M stars.
Such discrepancies between the data and simulations for M stars are not 
surprising because these stars are notoriously difficult to model due to 
their complicated chemistry (Allard {\em et al.} 1997). Particularly problematic 
are incomplete and uncertain opacity tables, as well as the possibility of 
dust condensation in the outer cool parts of the atmosphere, an effect which 
is usually ignored.

\section{   Conclusions     } 

This preliminary study indicates the enormous potential of combined 2MASS 
and SDSS data for stellar and Galactic structure studies.
The results of the positional matching of sources presented here 
indicate that both surveys will provide outstanding astrometry 
($\about 0.1$ arcsec rms per coordinate) for an unprecedented number of 
objects. 
Eight-color accurate photometry (better than 0.1 mag) for millions of stars 
is bound to place studies of  stellar and Galactic structure and evolution 
at an entirely new level. Such a large number of stars will 
enable detailed studies of the stellar initial mass function 
and its possible Galactic variation, of the metallicity gradients 
in the Galaxy, and similar other types of analyses which were until now 
impeded by the lack of large and reliable data sets. 
For example, by analyzing sources from less than 1\% of the final
2MASS-SDSS overlapping area, we find that either the parameters of the 
standard Galactic structure model, or the conventional initial mass 
function, must be changed to make simulations fit the data. Also, the 
wide and detailed spectral coverage based on eight fluxes provides strong 
tests for the stellar structure and atmosphere models. As we have shown here, 
the near-infrared model colors for M dwarfs are significantly different 
from the observed colors.

In addition to the testing of stellar and Galactic structure models as
discussed here, such a large and uniform data set is ideal for discovering
objects with unusual colors (e.g. T Tau stars, Herbig Ae/Be stars, brown
dwarfs, etc.). An analysis of such outliers in color-color space will
be presented in a forthcoming publication.

\vskip 0.4in
\leftline{Acknowledgments}

We are grateful to Martin Cohen and Mike Skrutskie for providing to us 
2MASS transmission curve data, and for helpful discussions. 

This publication makes use of data products from the Two Micron All Sky
Survey, which is a joint project of the University of Massachusetts and
the Infrared Processing and Analysis Center/California Institute of Technology, 
funded by the National Aeronautics and Space Administration and the National 
Science Foundation.

The Sloan Digital Sky Survey (SDSS) is a joint project of The University of
Chicago, Fermilab, the Institute for Advanced Study, the Japan Participation 
Group, The Johns Hopkins University, the Max-Planck-Institute for Astronomy, 
Princeton University, the United States Naval Observatory, and the University of 
Washington. Apache Point Observatory, site of the SDSS, is operated by the 
Astrophysical Research Consortium. Funding for the project has been provided by 
the Alfred P. Sloan Foundation, the SDSS member institutions, the National 
Aeronautics and Space Administration, the National Science Foundation, the U.S. 
Department of Energy and Monbusho, Japan. 
The SDSS Web site is http://www.sdss.org/.


\newpage

\begin{figure}
\label{SDSSall}
\plotfiddle{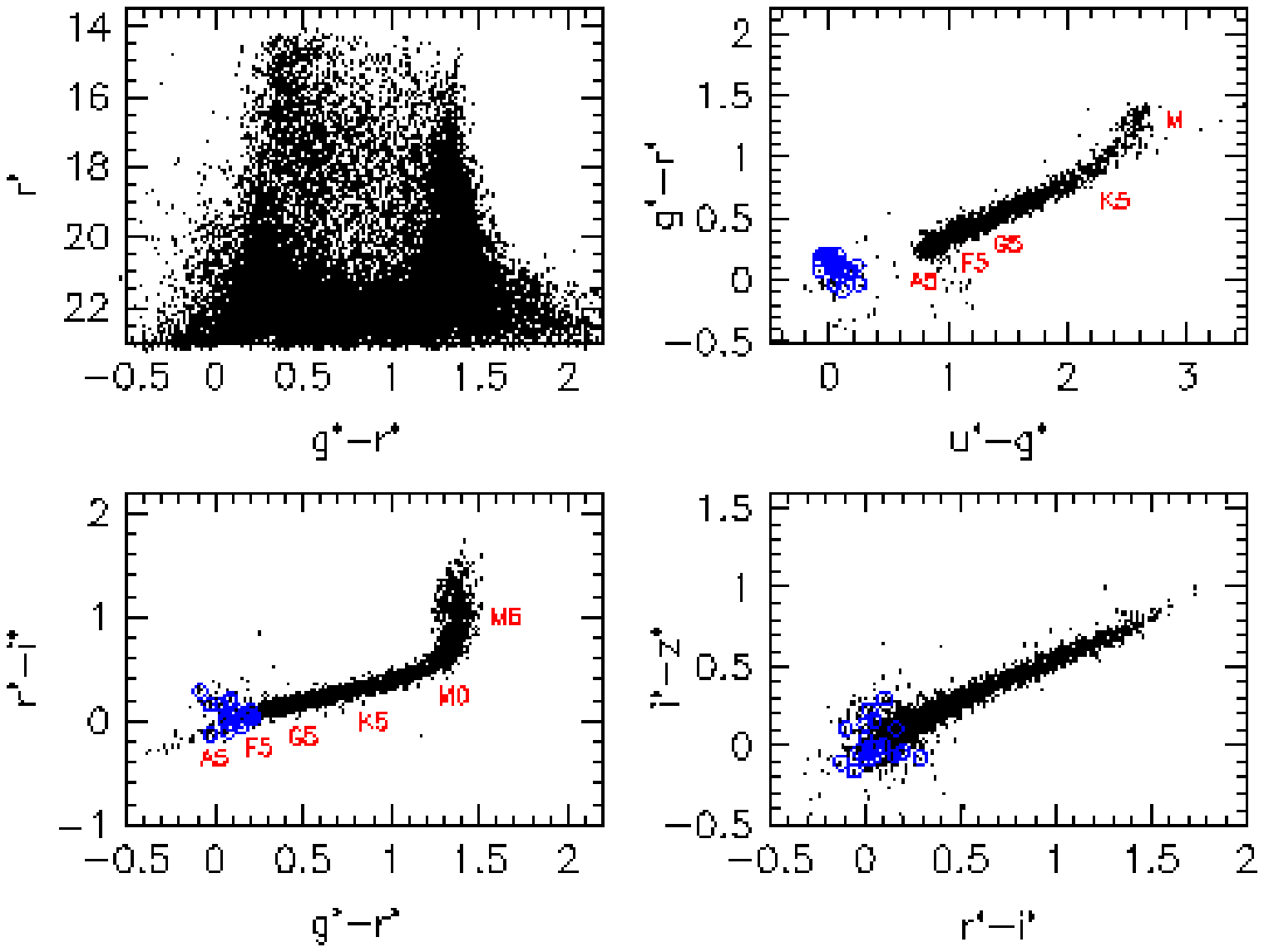}{12cm}{0}{100}{100}{-310}{-260}
\caption{The SDSS color-color and color-magnitude diagrams which summarize 
photometric properties of unresolved sources, marked as dots. The top left panel 
displays \r\ vs. \gr\ color-magnitude diagram for $\about$ 25,000 objects observed 
in 3 deg$^2$ of sky during SDSS commissioning run 94. The three remaining panels show 
color-color diagrams for objects brighter than 20$^m$ in each of the 3 bands used 
to construct each diagram (red is always towards the upper right corner). 
The locus of ``normal" stars is clearly visible in all three diagrams, 
and the positions of several spectral types are indicated next to the locus in 
the \ug\ vs. \gr\ and \ri\ vs. \gr\ color-color diagrams. Objects that have \ug\ 
and \gr\ colors similar to low-redshift QSOs (\ug $<$ 0.4, -0.1 $<$ \gr $<$ 0.3, 
\ri $<$ 0.5), and are brighter than the limit for quasars in the SDSS spectroscopic 
survey (\i $<$ 19), are marked by open circles.}    
\end{figure}

\begin{figure}
\label{RADec}
\plotfiddle{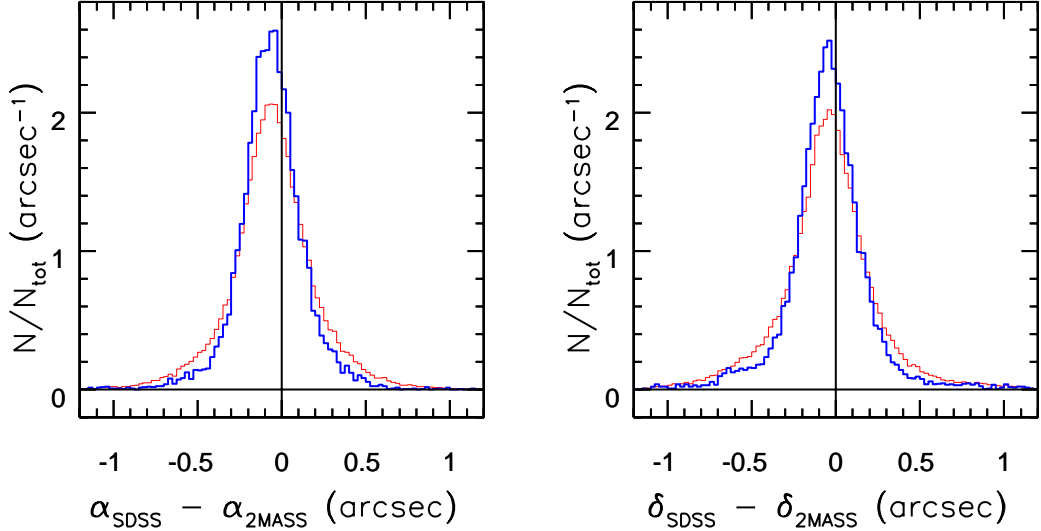}{6cm}{0}{75}{75}{-230}{-250}
\caption{The probability distributions of RA and Dec differences between 
the SDSS and 2MASS positions. The thin lines show the normalized histograms 
for all matched sources ($\about$ 63,000), and the thick lines show 
histograms for $\about$ 14,000 sources with the highest signal-to-noise 
ratios (see text). The equivalent Gaussian widths determined from the 
interquartile range are 0.30 arcsec for the distributions of all sources, and
0.15 arcsec for the high SNR sources. Note that the histograms show a 0.05 arcsec 
offset from the origin in both coordinates. This appears to be a systematic effect 
in the SDSS astrometry (see text).}
\end{figure}

\begin{figure}
\label{filters}
\plotfiddle{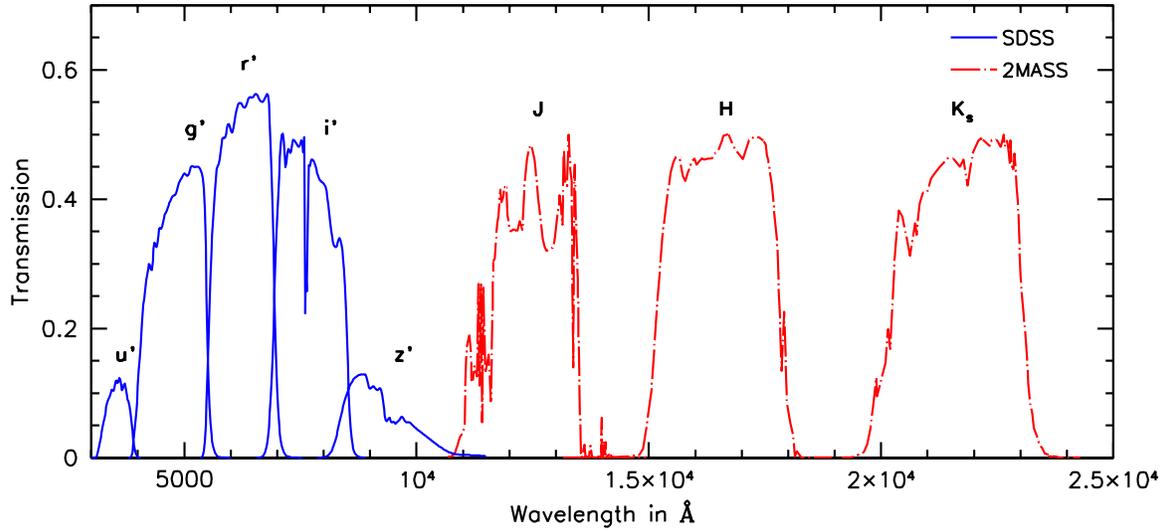}{5cm}{0}{80}{80}{-250}{-300}
\caption{The SDSS and 2MASS bandpasses. The five solid curves correspond to the
overall SDSS efficiencies and the three dot-dashed lines show the 2MASS 
responses normalized to 0.5 at the peak value. All curves include typical 
atmospheric extinction and the transmission of the entire instrumental system.}
\end{figure}

\begin{figure}
\label{cmd}
\plotfiddle{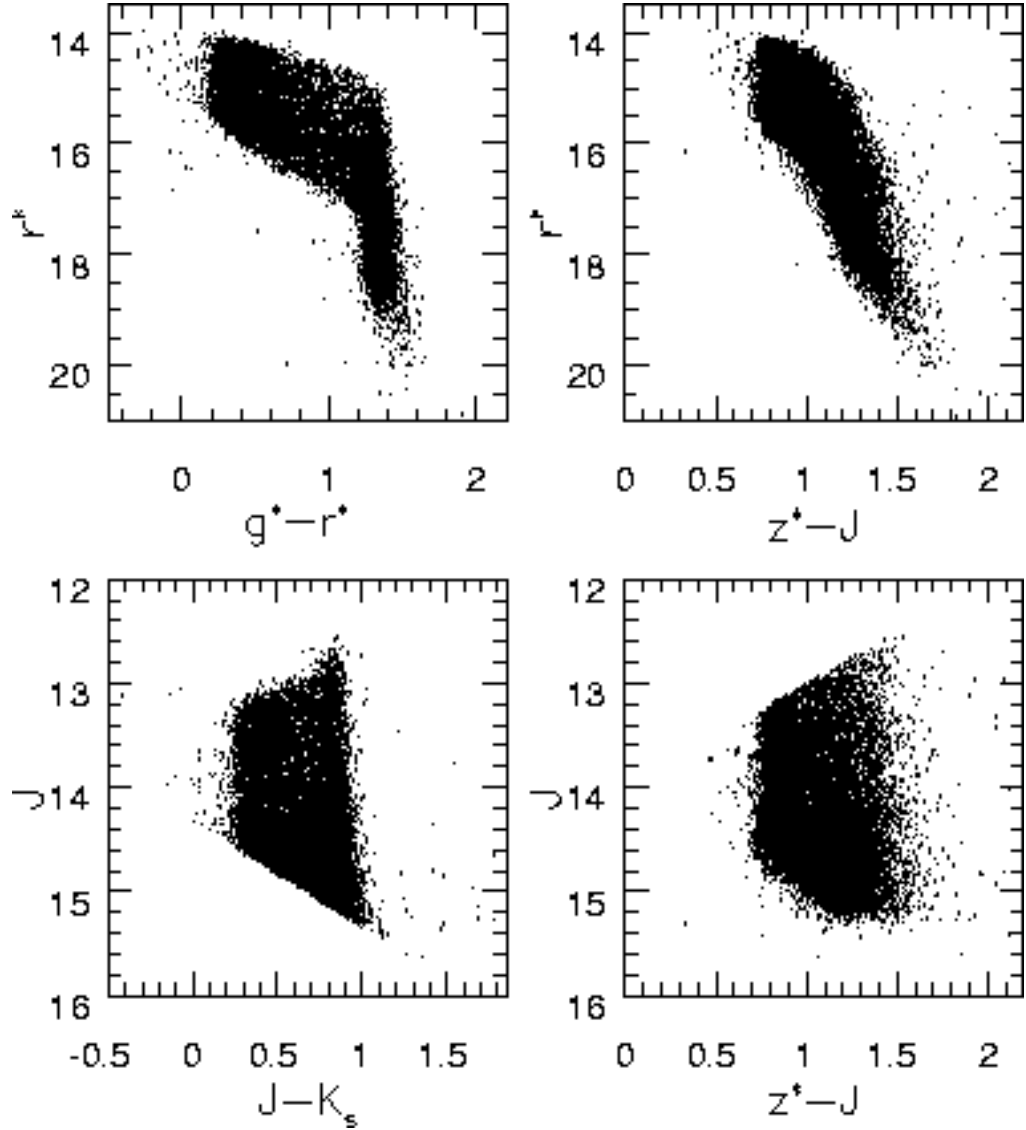}{14cm}{0}{100}{100}{-300}{-260}
\caption{The color-magnitude diagrams for $\about 14,000$ stars with the smallest 
($\la$ 10\%) photometric errors in both surveys. The bright limit is determined by 
the saturation in SDSS data, and the faint limit is determined by the 2MASS sensitivity.} 
\end{figure}

\begin{figure}
\label{ccd}
\plotfiddle{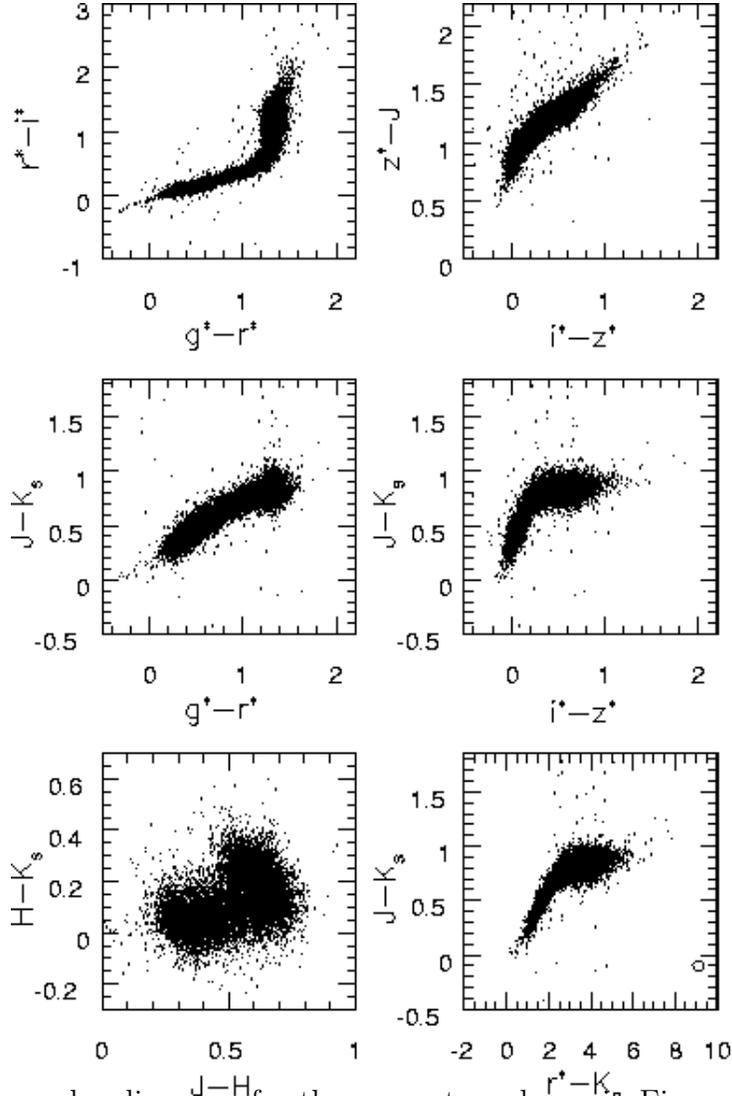}{12cm}{0}{70}{70}{-220}{-100}
\caption{The color-color diagrams for the same stars shown in Figure 4.  
The stellar locus is clearly visible in all diagrams that include at least
one optical band. In the near-infrared color-color diagram, shown in the lower 
left panel, the color ranges are much smaller, and consequently the stellar locus 
is not as well defined as in optical and optical-infrared color-color diagrams.
The position of Gl 229B, the first discovered T dwarf (Nakajima 
{\em et al.} 1995), is marked by an open circle at $r^*-K_s = 9.1$ and
$J-K_s = -0.1$ in the lower right panel.}    
\end{figure}

\begin{figure}
\label{contours}
\plotfiddle{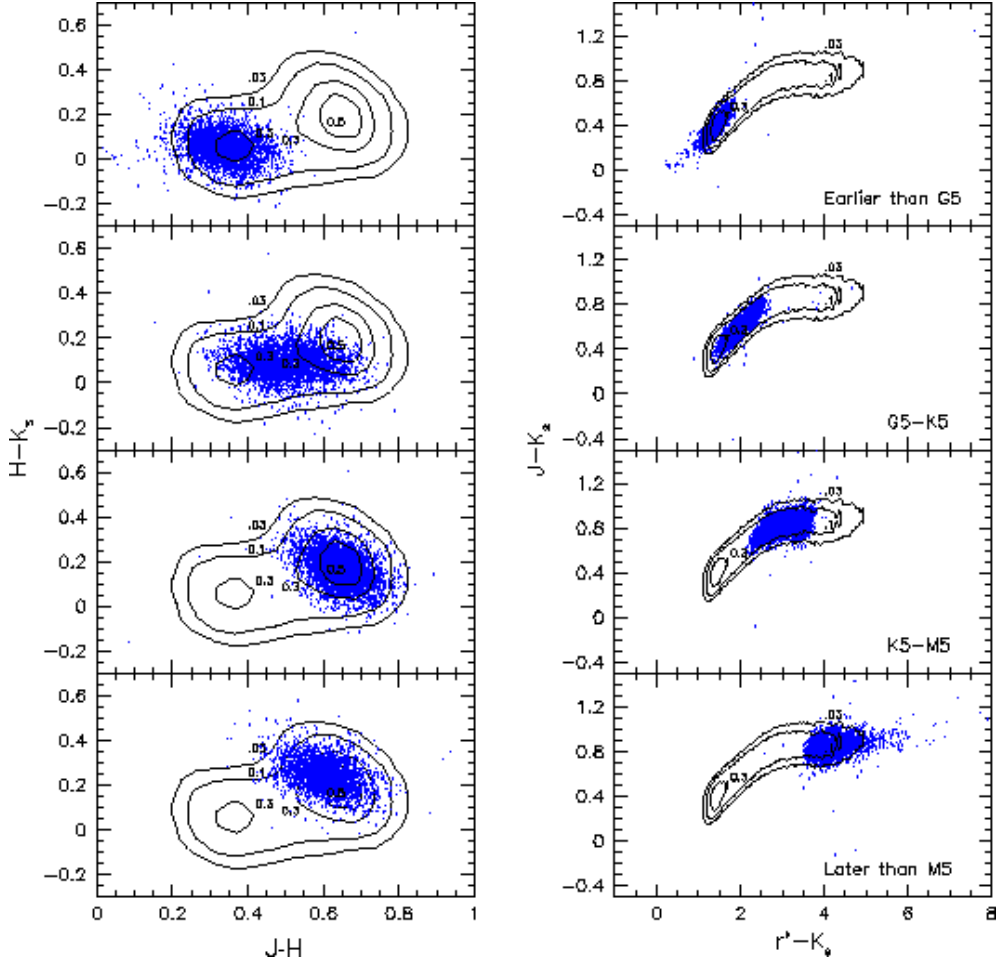}{14cm}{0}{70}{70}{-220}{-100}
\caption{The distribution of stars in four optically selected subsamples 
which trace the stellar spectral sequence in the $H-K_s$ vs. $J-H$ and $J-K_s$ 
vs. $r^*-K_s$  color-color diagrams. Stars in each subsample are shown as dots,
and the distribution of the whole sample is shown by contours. The contour levels
are 0.5, 0.3, 0.1 and 0.03 of the peak density. The top row includes stars 
with spectral types earlier than G5 (21\% of the whole sample), the second row
corresponds to spectral types between G5 and K5 (27\%), the third row includes
spectral types between K5 and M5 (32\%), and the bottom row includes stars 
later than M5 (21\%).}    
\end{figure}

\begin{figure}
\label{model_and_data_cc}
\plotfiddle{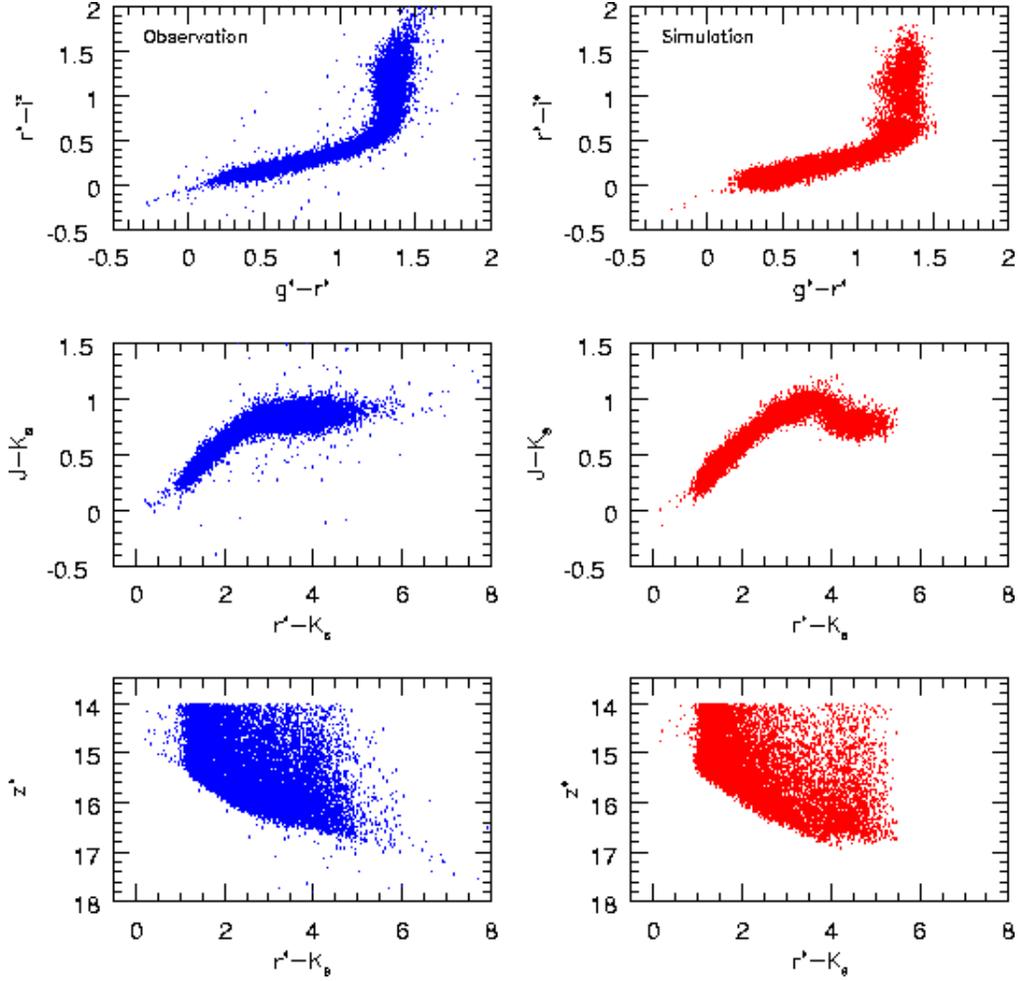}{14cm}{0}{70}{70}{-220}{-100}
\caption{Comparison of data with the predictions of a stellar population synthesis 
code. The panels on the left show the diagrams for the matched objects, and those 
on the right show the simulated population. The top row compares optical colors,
and the middle row shows an optical-infrared color-color diagram. The bottom row
compares the $z^*$ vs. $r^*-K_s$ color-magnitude diagrams.}    
\end{figure}

\begin{figure}
\label{matched_histograms}
\plotfiddle{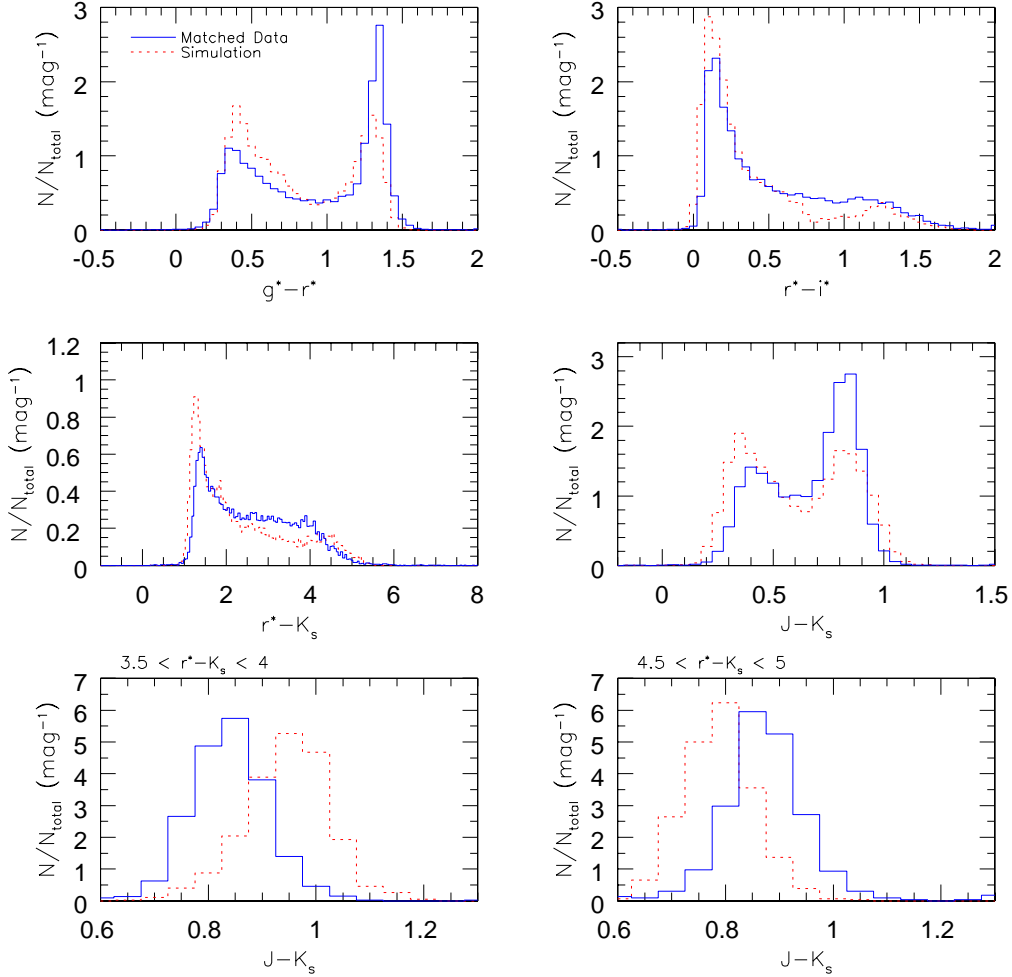}{14cm}{0}{70}{70}{-220}{-100}
\caption{Various color probability distributions for the matched and simulated 
objects shown in Figure 7. The top four panels show that the simulation
overestimates the overall counts of blue stars while underestimating the
redder star counts. The bottom two panels show the $J-K_s$ color distributions 
for two subsamples of M stars: the left panel shows stars with
$3.5 < \r -K_s < 4$ ($\about$ M3) and the right panel shows stars with
$4.5 < \r -K_s < 5$ ($\about$ M5). Note that the $J-K_s$ colors
for simulated early-M stars are significantly redder (0.15$^m$) than observed,
while they are bluer than observed for the mid-to-late-M stars.}    
\end{figure}

\end{document}